\documentclass[runningheads,a4paper]{llncs}
\usepackage{graphicx}
\usepackage{url}
\usepackage{float}
\usepackage{amsmath}
\usepackage{subfig}
\usepackage{dblfloatfix}
\usepackage[table]{xcolor}
\usepackage{color, colortbl}
\definecolor{Gray}{gray}{0.9}
\usepackage{graphicx,dblfloatfix}
\usepackage{tikz}
\usepackage{lipsum,afterpage}
\usetikzlibrary{trees,shapes}
\usetikzlibrary{decorations.pathmorphing}
\usetikzlibrary{decorations.markings}
\usetikzlibrary{arrows}
\usepackage{url}
\urldef{\mailsa}\path||
\urldef{\mailsa}\path| |
\urldef{\mailsc}\path||    
\newcommand{\keywords}[1]{\par\addvspace\baselineskip
\noindent\keywordname\enspace\ignorespaces#1}

\begin{document}

\mainmatter  


\title{An Empirical Study on Predictability of Software Code Smell Using Deep Learning Models}
\titlerunning{Software Code Smell Prediction Using Deep Learning Models}

\author{Himanshu Gupta\inst{1}\thanks{The work was done when authors were students in BITS Pilani, Hyderabad Campus.} 
\and Tanmay G. Kulkarni\inst{1}*
\and Lov Kumar\inst{1}  \and
Lalita Bhanu Murthy Neti\inst{1} \and
Aneesh Krishna\inst{2}
}
\authorrunning{H. Gupta et al.}

\institute{
	BITS Pilani, Hyderabad Campus, India 
	\email{ \{f20150339h,f20150647h\}@alumni.bits-pilani.ac.in}
	\email{ \{lovkumar,bhanu\}@hyderabad.bits-pilani.ac.in} 
	\and Curtin	University, Australia \\
	\email{a.krishna@curtin.edu.au}
}

\toctitle{Lecture Notes in Computer Science}
\tocauthor{Authors' Instructions}

\maketitle
\begin{abstract}
Code Smell, similar to a bad smell, is a surface indication of something tainted but in terms of software writing practices. This metric is an indication of a deeper problem lies within the code and is associated with an issue which is prominent to experienced software developers with acceptable coding practices. Recent studies have often observed that codes having code smells are often prone to a higher probability of change in the software development cycle. In this paper, we developed code smell prediction models with the help of features extracted from source code to predict eight types of code smell. Our work also presents the application of data sampling techniques to handle class imbalance problem and feature selection techniques to find relevant feature sets. Previous studies had made use of techniques such as Naive - Bayes and Random forest but had not explored deep learning methods to predict code smell. A total of 576 distinct Deep Learning models were trained using the features and datasets mentioned above. The study concluded that the deep learning models which used data from Synthetic Minority Oversampling Technique gave better results in terms of accuracy, AUC with the accuracy of some models improving from 88.47 to 96.84.

\keywords{Code Smell, Data Sampling Techniques, Software metrics, Feature selection}
\end{abstract}

\vspace*{-\baselineskip}

\vspace*{-\baselineskip}
\section{Introduction}
\label{intro}

A code smell is a quantifiable metric which indicates severe problems in complex software development life cycles due to poor programming practices. A code smell by itself may not reflect a programmatic error \cite{azeem2019machine,yamashita2012code} within the software. Instead, it is a harbinger of potential problems in the future during maintenance or when additional functionality is built into the software \cite{khomh2009exploratory}. A code smell is generally detected by inspecting the source code and searching for sections of the code can be restructured to improve the quality of code. This method is inefficient, especially if developers have to crawl through potentially thousands of lines of code, which can consume a significant amount of time and money to the organization. Based on the internal organization and anatomy of the software, a robust model can be created, which can make this excruciating process a lot simpler. In this work, we have used a set of metrics extracted from the source code of the software as an input to develop multiple models for predicting code smell present in the source code of the software. Time conserving ability and capability to maintain software will be improved if these hidden problems become apparent to developers.

The above-computed metrics are used as an input of the code smell prediction models, so the predictive ability of the models depends on the selection of relevant metrics. In our research, different methods were applied to find the relevant metrics and also methods that help to find these metrics. We observed that the considered datasets to validate this proposed work were highly imbalanced in terms of the number of samples. In order to balance them, we used different smoothing techniques. The primary focus of our work was to evaluate how different smoothing methods and different metric selection method affect the performance of a code smell prediction models.

After successful computation of the above steps, we have used different varieties of deep learning techniques to train the code smell, prediction models. We apply this model on the following code smells- Blob Class (BLOB), Complex Class (CC), Internal Getter/Setter (IGS), Leaking Inner Class (LIC), Long Method (LM), No Low Memory Resolver (NLMR), Member Ignoring Method (MIM), and Swiss Army Knife (SAK) \cite{hecht2016empirical}.  In this paper, we attempt to answer the following Research Questions (RQ):

\begin{itemize}
	\item \textbf{RQ1: Discuss the ability of selected features over the original features towards detecting Code Smell.}
	As there were many metrics to select from which the models were to be developed; it was inevitable that some of them were found to be co-related. We aim to determine the features which were related to predicted code smell. We also made sure that the selected features were unrelated to each other. The correlation between the features and resultant code smells were obtained using the Cross-Correlation Analysis and Wilcoxon Sign Rank test \cite{wilcoxon1970critical,podobnik2008detrended}.
	\item \textbf{RQ2: Discuss the ability of Data Sampling Techniques to detect Code Smell.}
	As the number of code smell and the metrics associated with them varies significantly with source code of the software, There was a need to sample them for creation of an unbiased dataset. This practice ensured that the models which were developed had balanced data to ensure proper training. ADASYN (Adaptive Synthetic Sampling Method) \cite{he2008adasyn} and SMOTE (Synthetic Minority Over-Sampling Method) \cite{bowyer2011smote} were used to balance the data.  
	\item \textbf{RQ3: Discuss the ability of different deep learning architectures to detect Code Smell.}
    Eight different deep learning architectures were developed in order to encounter the prediction of code smell. These neural networks were created by varying the number of hidden layers. The performance of these architectures was compared with accuracy, F-Measure and area under the curve.
\end{itemize}

\noindent \textbf{Organization:} The paper is organized as follows: Section \ref{sec:related} summarizes the related work. Section \ref{sec:research_background} gives a detailed overview of all the components used in the experiment. Section \ref{sec:research_framework} describes the research framework pipeline and how different components described in section \ref{sec:research_background} work together.Section \ref{sec:results} gives the experimental results and Section \ref{sec:comparison} answers the questions asked in Introduction. Finally, we conclude in Section \ref{sec:conclusion}.

\section{Related Work}
\label{sec:related}

Several studies were done for code smell detection and how they affect the performance of the system.
Yamashita et al. \cite{yamashita2012code} discuses an empirical method which recognises which code smells were recognised as significant for maintainability. The outcome is based on an analysis of an industrial case study done by the author.
Khomh et al. \cite{khomh2009exploratory} explores the connection between code smells and changes proneness in classes. The author has done intensive research to identify whether classes with code smells are more prone to change as compared to other classes and vice versa. The conclusion indicates that specific code smells adversely impact the classes.
Coleman et al. \cite{coleman1994using} demonstrate how software-related decision making can benefit from automated software maintainability analysis. 
Defect prediction using the Naïve Bayes method has been demonstrated by Wang et al. \cite{wang2010naive}. The author also analyses the construction of prediction models.
Turhan et al. \cite{turhan2007software} attempt to develop an approach that allows the use of software metrics to assist in defect prediction. The proposed methodology yields statistically better results.
Francesca Arcelli Fontana et al. \cite{fontana2016comparing} compared 16 different machine learning algorithms on four code smells. He found out that Random Forest performed the best while SVM performed the worst. 

There have been several studies for feature selection methods in an unbalanced dataset. Lida Abidi et al. \cite{abdi2015combat} presented an oversampling technique which generates synthetic samples of data using Mahalanobis distance which preserves the covariance of the minority distance.

Our primary contribution is to predict code smells utilizes deep learning in addition to data sampling techniques such as SMOTE and ADASYN to balance the dataset. We further make use of feature selection techniques to improve the accuracy of the models.

\section{Research Background}
\label{sec:research_background}
Our research is formulated using the following steps :
\begin{itemize}
\item In order to encounter the data imbalance problem, we created two different datasets apart from the original.  
\item We then select the relevant features from each dataset via cross co-relation analysis and Wilcoxon Sign Rank test.
\item After normalizing the datasets, we train eight deep learning models on them and validate them using 5-fold cross-validation.
\end{itemize}

\subsection{Experimental Dataset}
In this paper, we have analyzed the code-bases of 629 open source projects which were scraped from GitHub. This dataset consists of a list of packages along with the corresponding code smells. The Anti-Patterns observed over the repositories are given in Table \ref{tab1}. From the table, it is evident that each code smell varies greatly. Some were very common as to be present in 75.2\% of the entire dataset. At the same time, some of the smells were as scarce as being present in just 26.4\% of the dataset. Furthermore, we observe that the least commonly observed code smell is the Swiss Army Knife (SAK), and most commonly observed code smell is the Long Method (LM).

\begin{table}[t!]
	\centering
	\caption{Dataset description which give code smell numbers and percentage} \label{tab1}
	\begin{tabular}{|l|l|l|l|l|}
			\hline
			\textbf{\begin{tabular}[c]{@{}l@{}}Code Smell \\ Type\end{tabular}} & \textbf{\begin{tabular}[c]{@{}l@{}}No Smell\\ Number\end{tabular}} & \textbf{\begin{tabular}[c]{@{}l@{}}No Smell\\ Percentage\end{tabular}} & \textbf{\begin{tabular}[c]{@{}l@{}}Smelly\\ Number\end{tabular}} & \textbf{\begin{tabular}[c]{@{}l@{}}Smelly \\ Percent\end{tabular}} \\ \hline
			\textbf{BLOB}                                                       & 236                                                                & 37.59\%                                                                & 393                                                              & 62.48\%                                                            \\ \hline
			\textbf{LM}                                                         & 156                                                                & 24.80\%                                                                & 475                                                              & 75.20\%                                                            \\ \hline
			\textbf{SAK}                                                        & 463                                                                & 73.60\%                                                                & 166                                                              & 26.40\%                                                            \\ \hline
			\textbf{CC}                                                         & 188                                                                & 29.88\%                                                                & 441                                                              & 70.12\%                                                            \\ \hline
			\textbf{IGS}                                                        & 277                                                                & 44.03\%                                                                & 352                                                              & 55.97\%                                                            \\ \hline
			\textbf{MIM}                                                        & 261                                                                & 41.49\%                                                                & 368                                                              & 58.51\%                                                            \\ \hline
			\textbf{NLMR}                                                       & 158                                                                & 25.11\%                                                                & 471                                                              & 74.89\%                                                            \\ \hline
			\textbf{LIC}                                                        & 227                                                                & 36.08\%                                                                & 402                                                              & 63.92\%                                                            \\ \hline
		\end{tabular}
	\vspace{-0.5cm}
\end{table}

\subsection{Software Metrics}

\label{metrics}
Software metrics are quantitative measures of different aspects of code features of a software. The different aspects include features like the modularity of a class, size of the class and many other similar characteristics. Although these metrics are used in a variety of tasks related to software engineering like software performance, measuring productivity and other software engineering tasks, we utilized these metrics as features in code smell detection of the software. There were four primary types of software metrics used for this study :
\begin{itemize}
	\item \textbf{Dimensional Metrics:}
	These metrics are aimed towards providing a quantitative metric for understanding code sizes and modularity. One might believe that more code leads to more features, but it makes the code difficult to handle in the long run \cite{abd2012metrics}.
	\item \textbf{Complexity Metrics:}
	These metrics help us gauge the complexity of applications, which are associated with the fact that an increase in complexity makes it challenging to comprehend and thus, difficult to test and maintain due to a larger number of paths of execution \cite{fontana2016comparing}.
	\item \textbf{Object Oriented Metrics:}
	These metrics are used to find the complexity, cohesion and coupling between the software modules \cite{harrison1998evaluation}.
	\item \textbf{Android Oriented Metrics:}
	These metrics show how Android specific dependencies and operations affect execution speed and User Experience.
\end{itemize}

\subsection{Data Sampling Techniques to Handle Imbalanced Data}

As the distribution of the target classes, as shown in Table \ref{tab1} is skewed within the dataset. Thus, we make use of the following standard data sampling techniques to offset the probability of each class in the dataset: 
\begin{itemize}
    \item ADASYN \cite{he2008adasyn} generates additional data points by using the predefined density distribution of the dataset.
    \item SMOTE \cite{bowyer2011smote} creates additional data points by assuming a uniform distribution of the dataset.
\end{itemize}

\subsection{Feature Selection Techniques}
In order to find the relevant code smell metrics, we make use of the Wilcoxon Sign Rank Test \cite{wilcoxon1970critical} and Cross-Correlation analysis \cite{podobnik2008detrended}. Using the methods mentioned above was important our research as we wanted to use only the statistically significant (highly unrelated to each other and related to output variable) metrics for creation of the models. After we find the significant features with the help of these techniques, we apply cross-correlation analysis to find not related features. For the above premise to work efficiently, the following null hypothesis has been proposed: Feature metric is incapable of finding out a particular code smell. In this study, we have made use of a p-value of 0.05; that is, we reject the hypothesis if the probability of the null hypothesis is below 0.05.

\subsection{Deep Learning Model for Classification}
In this study, eight deep learning models were used to train the models for predicting different types of code smell. Figures \ref{model1} and \ref{model2} show the architecture of the considered deep learning model1 and model2 (DL1 and DL2). Similarly, we are increasing the number of hidden layers for six more deep learning models (DL3: 3 hidden layers, DL4: 4 hidden layers, DL5: 5 hidden layers, DL6: 6 hidden layers, DL7: 7 hidden layers, DL8: 8 hidden layers). The following are the details of the neural networks shown in Figures \ref{model1} and \ref{model2} :

\begin{figure}[h]
\renewcommand{\thesubfigure}{\thefigure.\arabic{subfigure}}
	\centering
	\subfloat[1 hidden layer\label{model1}]
	{
		\includegraphics[width=5.0cm, height= 4.50cm]{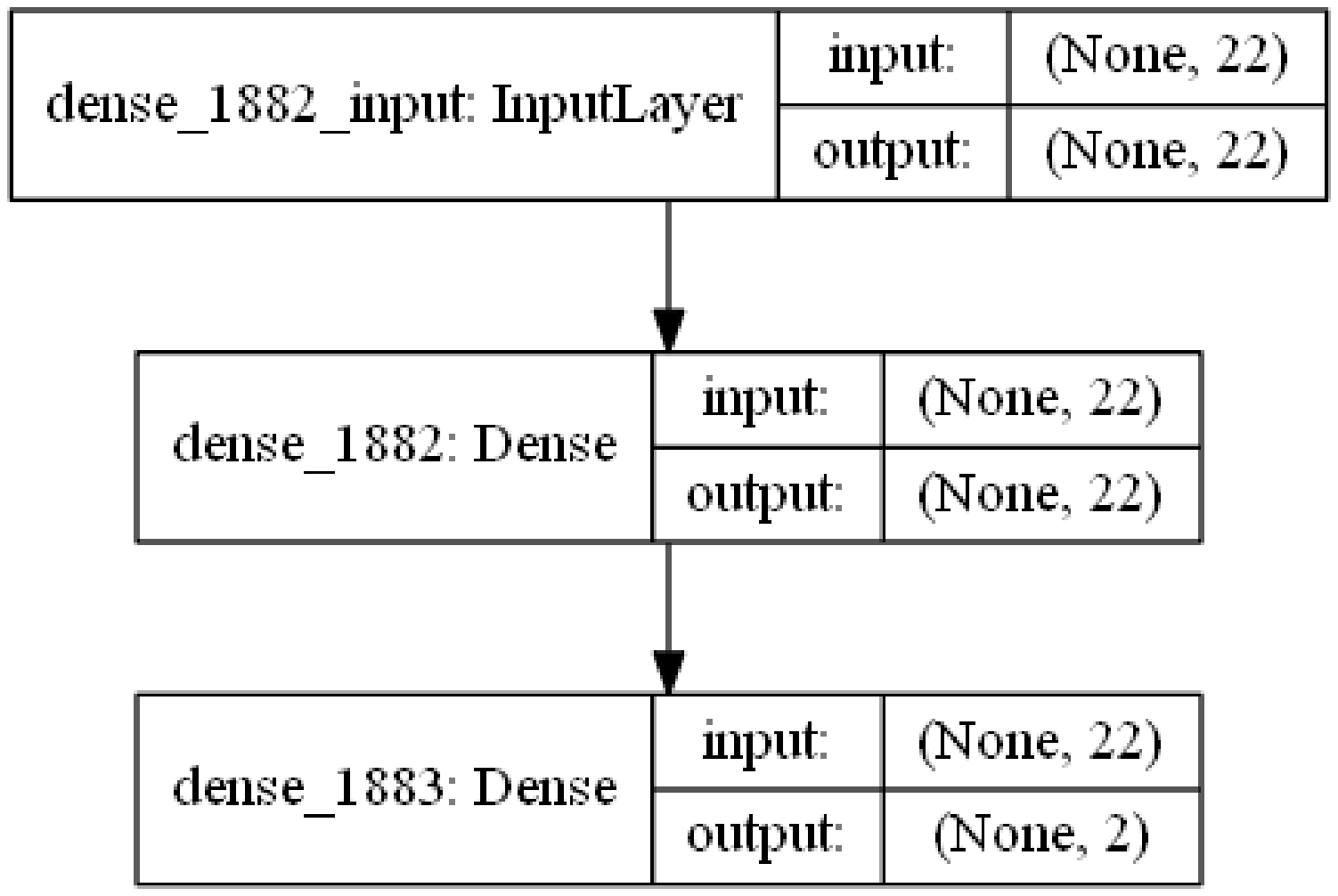}
	} 
	\subfloat[2 hidden layers\label{model2}]
	{
		\includegraphics[width=5.0cm, height=4.0cm]{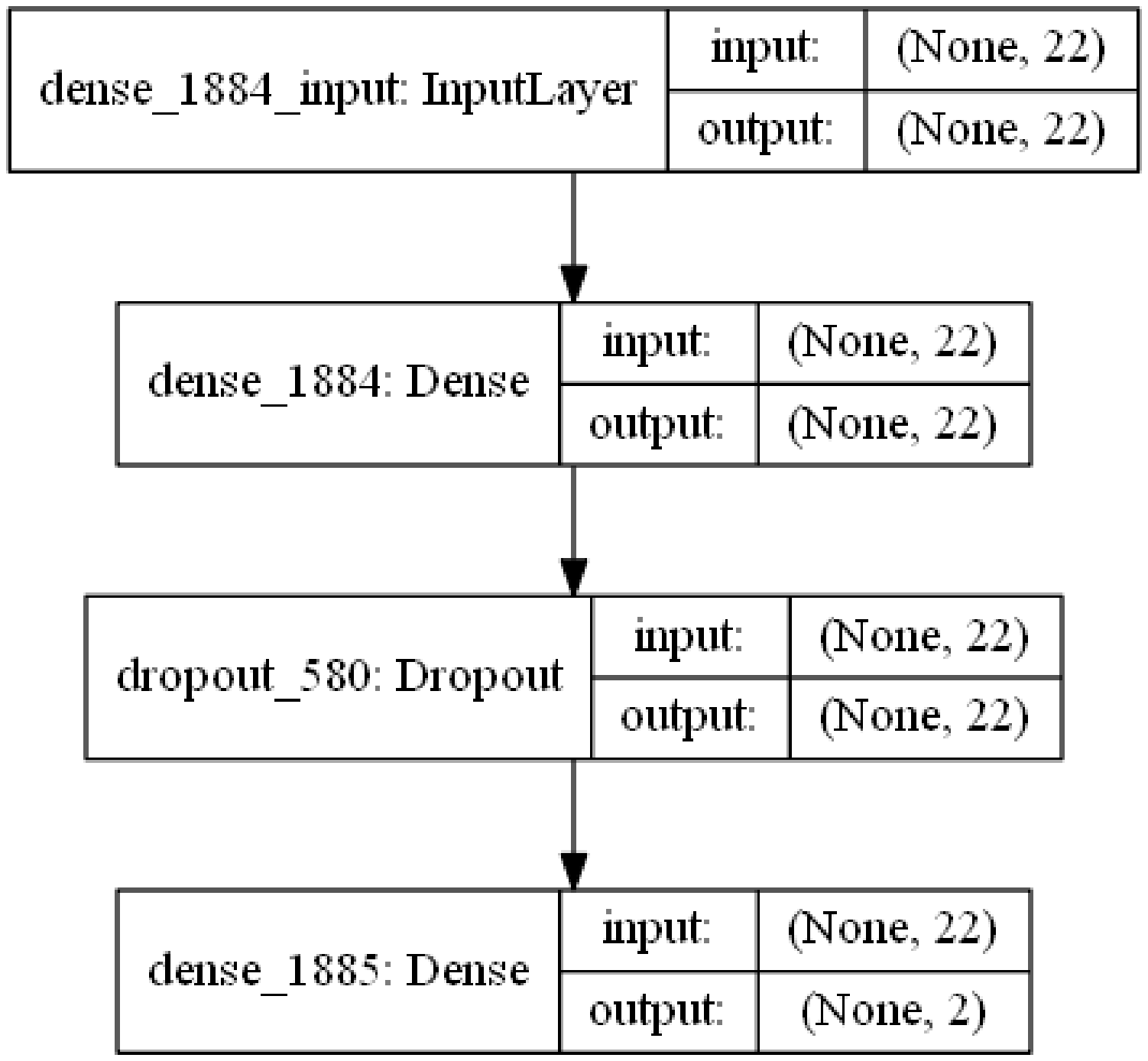} 
	} 
	\label{modeln}
	\caption[Deep Learning Model Framework]%
    {\centering Deep Learning Model Framework \par \small \textbf{Note:} \textit{None} in the figure refers to the Batch size which is flexible. }
    \vspace{-0.5cm}
\end{figure}

\begin{itemize}
	\item The model has an input layer with 22 nodes corresponding to 22 features given as an input to the model.
	\item  It is followed by a hidden layer with 22 nodes and ReLU \cite{agarap2018deep} function is used as an activation function.
	\item The output layer has 2 nodes with softmax \cite{wang2018high} as the activation function.
	\item For the 2nd model we have added a dropout layer to prevent over-fitting with a probability of 0.2 after the hidden layer.
	\item Rest of the 6 models follow a similar pattern by adding dense layer and a dropout serially one after another with ReLU being the activation function. 
	\item All the models are trained with a batch size of 32 and 100 epochs.
\end{itemize}

\section{Research Framework}
\label{sec:research_framework}

\begin{figure*}[b!]
    \vspace{-0.5cm}
	\centering
	\includegraphics[width=11cm, height= 3.5cm]{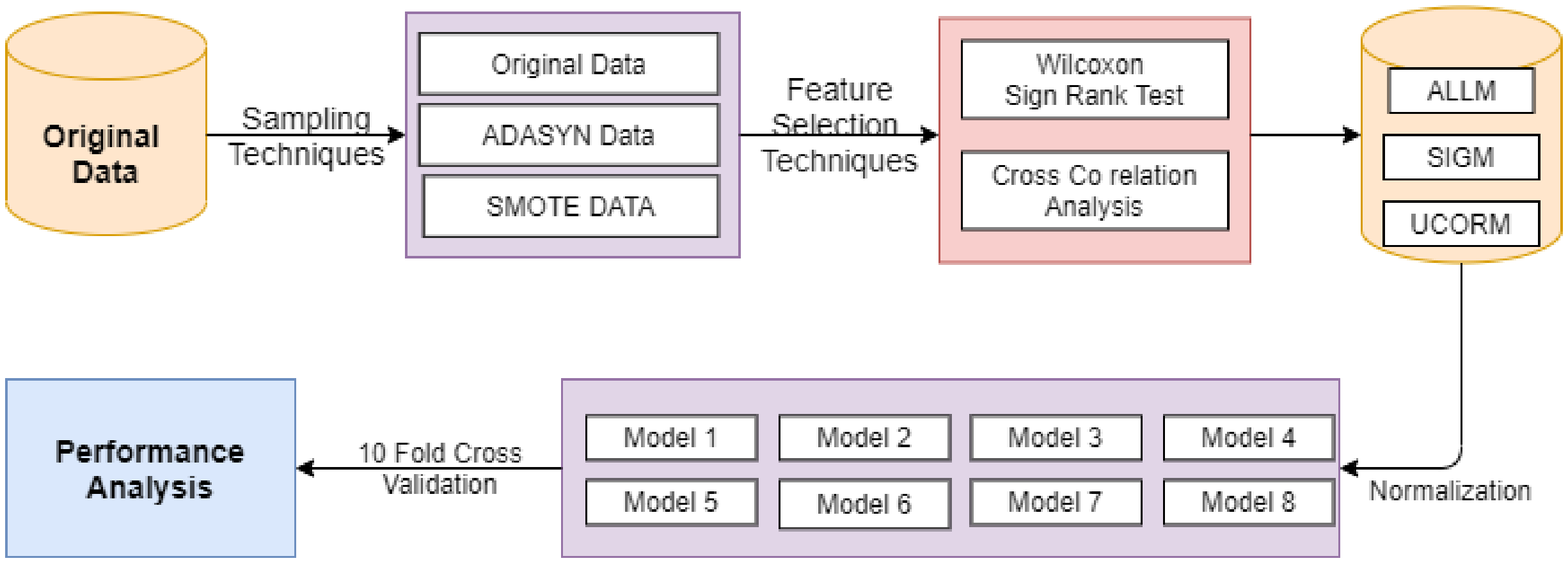}
	\caption{Flowchart of the Research Framework}
	\label{fig1}
	 \vspace{-1cm}
\end{figure*}

As shown in Figure \ref{fig1},  SMOTE and ADASYN sampling methods are applied to the dataset to tackle the class imbalance problem. Then, we have made use of statistical significance tests (Cross Co-relation and Wilcoxon Sign rank test) to identify the relevant metrics. These relevant metrics were used to train eight deep learning models and are compared using accuracy and AUC. We have also considered a significant test to validate the considered hypothesis: There is no significant improvement in the predictive ability of the models after applying data sampling techniques and feature selection techniques.

\begin{figure}[t!]
	\renewcommand{\thesubfigure}{\thefigure.\arabic{subfigure}}
	\centering
	\subfloat[BLOB\label{corr1}]{
		\includegraphics[width=3.5cm, height=3.5cm]{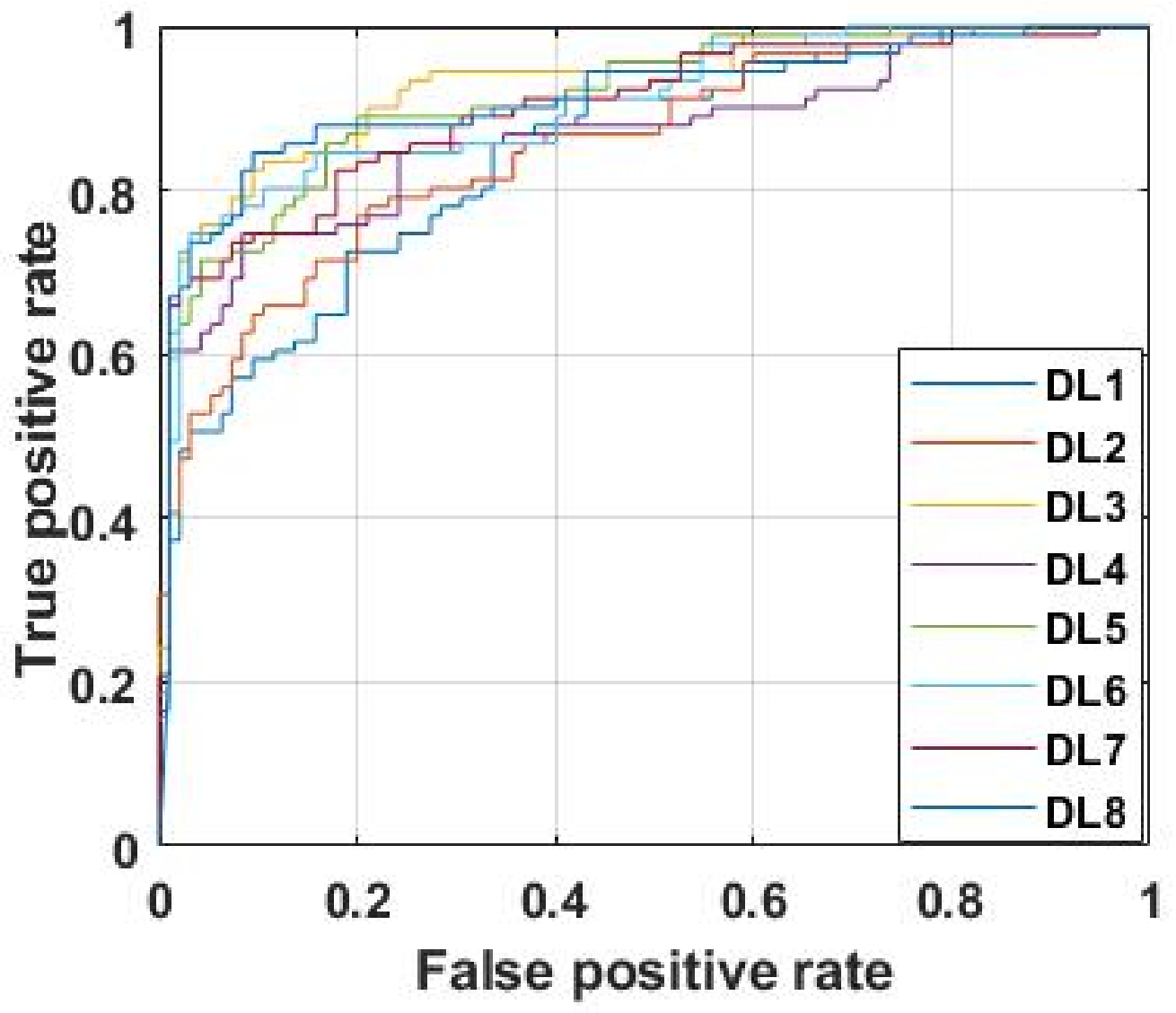} }
	\subfloat[LK\label{corr2}]{
		\includegraphics[width=3.5cm, height=3.5cm]{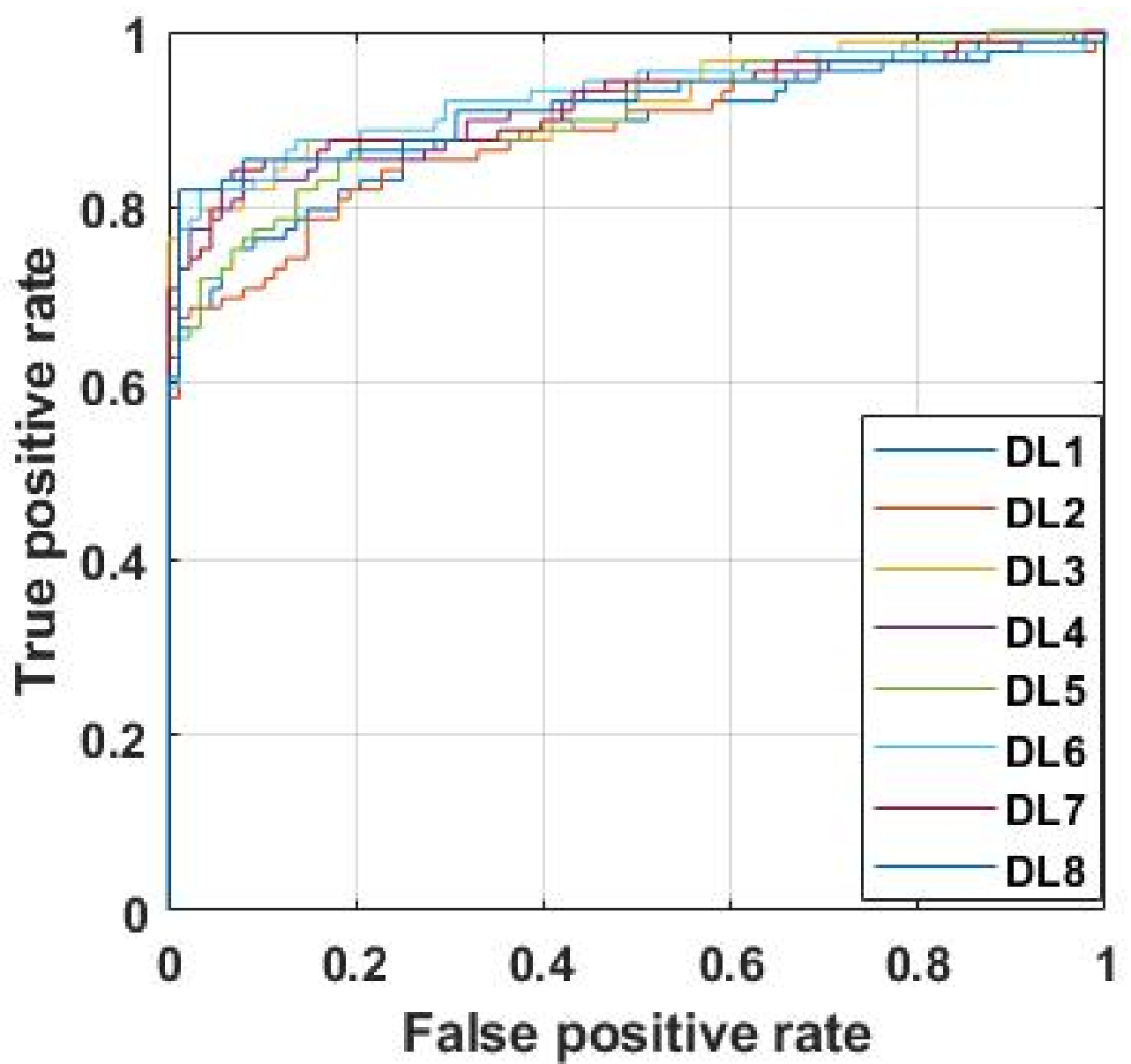} } 
	\subfloat[SAK\label{corr3}]{
		\includegraphics[width=3.5cm, height=3.5cm]{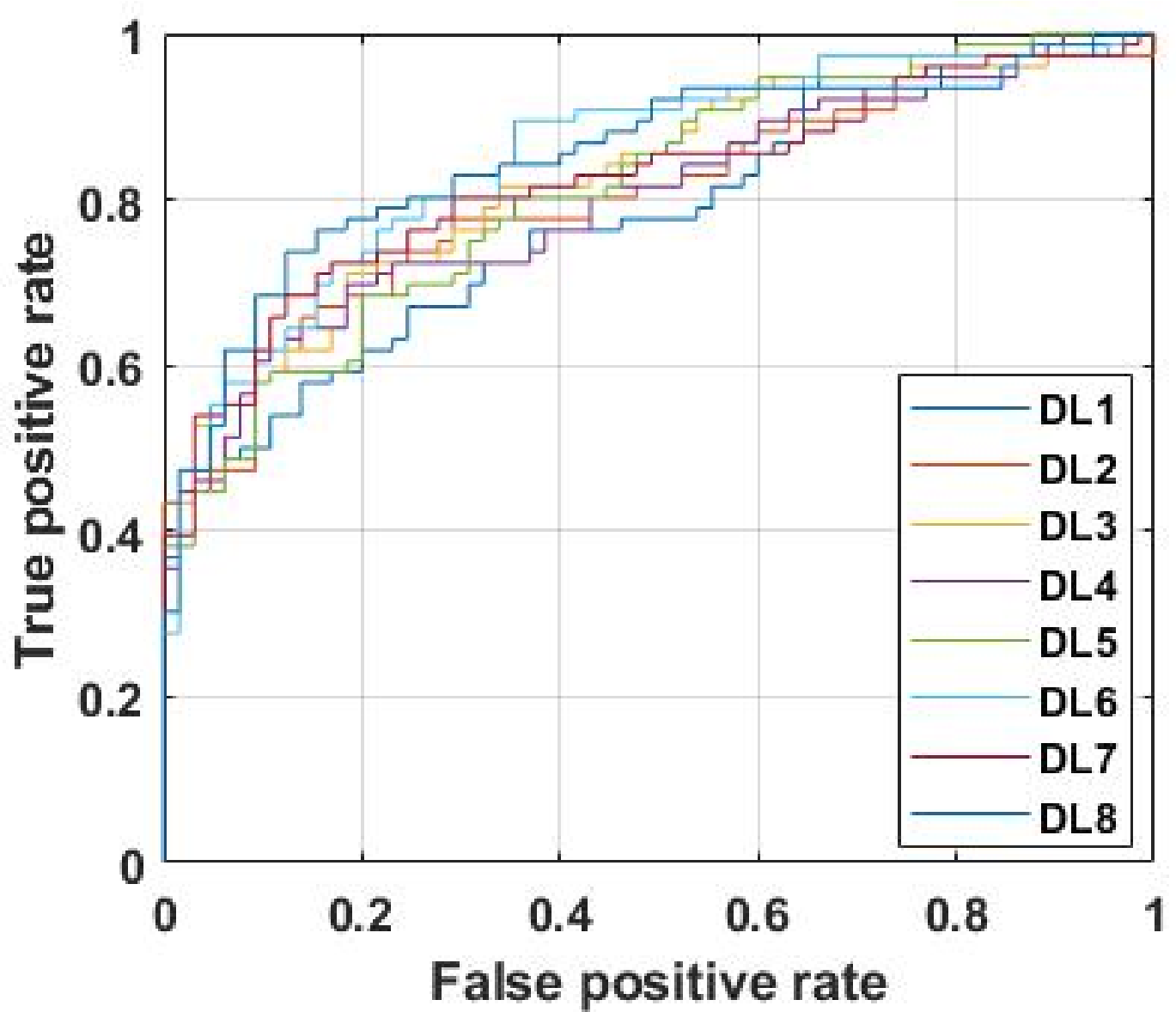} }
	\caption{ROC curve: BLOB, LM. and SAK prediction models}
	\label{conf}
	 \vspace{-0.5cm}
\end{figure}

\begin{table*}[b!]
    \vspace{-1cm}
	\centering
	\caption{ Accuracy, and  AUC values for Deep learning prediction models trained on SMOTE data} \label{tab21}
	\renewcommand{\arraystretch}{1.1}
	\resizebox{11.0cm}{!}{
		\begin{tabular}{|l|c|c|c|c|c|c|c|c|c|c|c|c|c|c|c|c|c|}
			\hline
			& \multicolumn{8}{|c|}{\textbf{Accuracy}}   &       \multicolumn{8}{|c|}{\textbf{AUC}}                                                                                                                                                                                              \\ \hline

			\multicolumn{17}{|c|}{\textbf{	ALLM}} \\ \hline
			
			&\multicolumn{1}{l|}{\textbf{DL1}} & \multicolumn{1}{l|}{\textbf{DL2}} & \multicolumn{1}{l|}{\textbf{DL3}} & \multicolumn{1}{l|}{\textbf{DL4}} & \multicolumn{1}{l|}{\textbf{DL5}} & \multicolumn{1}{l|}{\textbf{DL6}} & \multicolumn{1}{l|}{\textbf{DL7}} & \multicolumn{1}{l|}{\textbf{DL8}} & \multicolumn{1}{l|}{\textbf{DL1}} & \multicolumn{1}{l|}{\textbf{DL2}} & \multicolumn{1}{l|}{\textbf{DL3}} & \multicolumn{1}{l|}{\textbf{DL4}} & \multicolumn{1}{l|}{\textbf{DL5}} & \multicolumn{1}{l|}{\textbf{DL6}} & \multicolumn{1}{l|}{\textbf{DL7}} & \multicolumn{1}{l|}{\textbf{DL8}} \\ \hline
			
			BLOB & 74.52 & 75.80 & 77.71 & 78.34 & 75.16 & 80.25 & 82.80 & \cellcolor{green!20}83.44 & 0.83 & 0.86 & 0.87 & \cellcolor{green!20}0.88 & 0.83 & 0.86 & 0.87 & \cellcolor{green!20}0.88 \\ \hline
			LM & 90.00 & 87.89 & 94.74 & 93.16 & 93.16 & \cellcolor{green!20}96.32 & 94.21 & \cellcolor{green!20}96.32 & 0.97 & 0.98 & \cellcolor{green!20}0.99 & \cellcolor{green!20}0.99 & 0.97 & 0.98 & \cellcolor{green!20}0.99 & \cellcolor{green!20}0.99 \\ \hline
			SAK & 76.34 & 77.96 & \cellcolor{green!20}84.95 & 79.03 & 82.26 & \cellcolor{green!20}84.95 & 79.57 & \cellcolor{green!20}84.95 & 0.85 & 0.90 & \cellcolor{green!20}0.91 & \cellcolor{green!20}0.91 & 0.85 & 0.90 & \cellcolor{green!20}0.91 & \cellcolor{green!20}0.91 \\ \hline
			CC & 82.49 & 80.79 & 87.01 & 86.44 & 83.62 & 88.14 & 88.14 & \cellcolor{green!20}89.27 & 0.89 & \cellcolor{green!20}0.92 & \cellcolor{green!20}0.92 & \cellcolor{green!20}0.92 & 0.89 & \cellcolor{green!20}0.92 & \cellcolor{green!20}0.92 & \cellcolor{green!20}0.92 \\ \hline
			IGS & 70.21 & 73.05 & 75.18 & 72.34 & 72.34 & 76.60 & 73.76 & \cellcolor{green!20}78.01 & 0.79 & 0.81 & 0.83 & \cellcolor{green!20}0.84 & 0.79 & 0.81 & 0.83 & \cellcolor{green!20}0.84 \\ \hline
			MIM & 74.32 & 71.62 & 78.38 & 81.08 & 75.00 & \cellcolor{green!20}82.43 & 73.65 & 80.41 & 0.83 & \cellcolor{green!20}0.89 & 0.87 & 0.86 & 0.83 & \cellcolor{green!20}0.89 & 0.87 & 0.86 \\ \hline
			NLMR & 89.95 & 89.42 & \cellcolor{green!20}94.71 & 91.53 & 91.01 & 92.59 & 93.12 & 94.18 & 0.96 & \cellcolor{green!20}0.98 & \cellcolor{green!20}0.98 & \cellcolor{green!20}0.98 & 0.96 & \cellcolor{green!20}0.98 & \cellcolor{green!20}0.98 & \cellcolor{green!20}0.98 \\ \hline
			LIC & 81.99 & 78.26 & 86.96 & 85.71 & 84.47 & 83.85 & 83.23 & \cellcolor{green!20}89.44 & 0.90 & 0.92 & 0.92 & \cellcolor{green!20}0.93 & 0.90 & 0.92 & 0.92 & \cellcolor{green!20}0.93 \\ \hline
			
			\multicolumn{17}{|c|}{\textbf{SIGM}} \\ \hline
			BLOB & 74.52 & 75.16 & 73.25 & 76.43 & 77.07 & 74.52 & 78.98 & \cellcolor{green!20}80.25 & 0.82 & 0.84 & 0.84 & \cellcolor{green!20}0.87 & 0.82 & 0.84 & 0.84 & \cellcolor{green!20}0.87 \\ \hline
			LM & 89.47 & 88.95 & 93.68 & 93.68 & 94.21 & \cellcolor{green!20}96.84 & 93.68 & 95.26 & 0.96 & \cellcolor{green!20}0.98 & \cellcolor{green!20}0.98 & \cellcolor{green!20}0.98 & 0.96 & \cellcolor{green!20}0.98 & \cellcolor{green!20}0.98 & \cellcolor{green!20}0.98 \\ \hline
			SAK & 63.98 & \cellcolor{green!20}66.67 & 65.05 & 60.22 & 62.37 & 64.52 & 62.90 & 61.83 & 0.67 & 0.71 & \cellcolor{green!20}0.74 & 0.64 & 0.67 & 0.71 & \cellcolor{green!20}0.74 & 0.64 \\ \hline
			CC & 80.79 & 82.49 & 85.31 & 83.05 & 84.18 & \cellcolor{green!20}88.7 & 87.01 & 87.01 & 0.88 & 0.90 & 0.92 & \cellcolor{green!20}0.93 & 0.88 & 0.90 & 0.92 & \cellcolor{green!20}0.93 \\ \hline
			IGS & 68.09 & 64.54 & \cellcolor{green!20}73.76 & 69.50 & 70.92 & 73.05 & 72.34 & \cellcolor{green!20}73.76 & 0.74 & 0.79 & 0.79 & \cellcolor{green!20}0.81 & 0.74 & 0.79 & 0.79 & \cellcolor{green!20}0.81 \\ \hline
			MIM & 71.62 & 73.65 & 76.35 & 70.95 & 70.95 & \cellcolor{green!20}79.73 & 72.30 & 78.38 & 0.80 & 0.83 & 0.83 & \cellcolor{green!20}0.85 & 0.80 & 0.83 & 0.83 & \cellcolor{green!20}0.85 \\ \hline
			NLMR & 89.42 & 88.89 & 92.59 & 88.89 & 93.65 & 94.18 & 92.59 & \cellcolor{green!20}94.71 & 0.96 & 0.97 & \cellcolor{green!20}0.98 & \cellcolor{green!20}0.98 & 0.96 & 0.97 & \cellcolor{green!20}0.98 & \cellcolor{green!20}0.98 \\ \hline
			LIC & 77.02 & 81.37 & 83.23 & 81.99 & \cellcolor{green!20}85.09 & 83.23 & 82.61 & \cellcolor{green!20}85.09 & 0.90 & 0.91 & \cellcolor{green!20}0.92 & \cellcolor{green!20}0.92 & 0.90 & 0.91 & \cellcolor{green!20}0.92 & \cellcolor{green!20}0.92 \\ \hline
			
			\multicolumn{17}{|c|}{\textbf{UCORM}} \\ \hline
			BLOB & \cellcolor{green!20}80.89 & 77.71 & 73.89 & 77.07 & 77.07 & 78.34 & 79.62 & 75.16 & \cellcolor{green!20}0.84 & \cellcolor{green!20}0.84 & \cellcolor{green!20}0.84 & 0.83 & \cellcolor{green!20}0.84 & \cellcolor{green!20}0.84 & \cellcolor{green!20}0.84 & 0.83 \\ \hline
			LM & 87.89 & 87.89 & 89.47 & 89.47 & 89.47 & 90.00 & 89.47 & \cellcolor{green!20}91.05 & 0.96 & 0.96 & 0.96 & \cellcolor{green!20}0.97 & 0.96 & 0.96 & 0.96 & \cellcolor{green!20}0.97 \\ \hline
			SAK & 64.52 & 63.98 & 63.44 & 63.98 & 65.05 & 66.13 & 65.59 & \cellcolor{green!20}70.97 & 0.70 & 0.71 & \cellcolor{green!20}0.73 & 0.69 & 0.70 & 0.71 & \cellcolor{green!20}0.73 & 0.69 \\ \hline
			CC & 81.36 & 80.23 & 84.18 & 81.36 & 82.49 & \cellcolor{green!20}86.44 & 79.66 & 83.62 & 0.89 & 0.89 & \cellcolor{green!20}0.9 & \cellcolor{green!20}0.9 & 0.89 & 0.89 & \cellcolor{green!20}0.9 & \cellcolor{green!20}0.9 \\ \hline
			IGS & 65.96 & 63.83 & 68.79 & 68.79 & 67.38 & \cellcolor{green!20}71.63 & 66.67 & 69.50 & 0.73 & 0.78 & 0.79 & \cellcolor{green!20}0.81 & 0.73 & 0.78 & 0.79 & \cellcolor{green!20}0.81 \\ \hline
			MIM & 70.95 & 70.95 & 70.95 & 71.62 & 70.95 & \cellcolor{green!20}73.65 & 70.95 & 71.62 & 0.81 & \cellcolor{green!20}0.84 & \cellcolor{green!20}0.84 & 0.82 & 0.81 & \cellcolor{green!20}0.84 & \cellcolor{green!20}0.84 & 0.82 \\ \hline
			NLMR & 88.89 & 89.42 & 88.36 & 88.36 & \cellcolor{green!20}90.48 & 88.36 & 89.95 & 87.83 & 0.95 & 0.95 & \cellcolor{green!20}0.96 & \cellcolor{green!20}0.96 & 0.95 & 0.95 & \cellcolor{green!20}0.96 & \cellcolor{green!20}0.96 \\ \hline
			LIC & 78.26 & 78.26 & 78.88 & 80.12 & 81.37 & 83.85 & \cellcolor{green!20}85.09 & 83.23 & 0.87 & 0.88 & 0.90 & \cellcolor{green!20}0.92 & 0.87 & 0.88 & 0.90 & \cellcolor{green!20}0.92 \\ \hline

	\end{tabular}}
\end{table*}

\section{Experiment Results}
\label{sec:results}
In this work, eight neural networks were used for developing a model to classify different types of code smell by considering extracted features from source code as an input whose details were presented in sub-section \ref{metrics}. SMOTE and ADASYN were used to tackle the class imbalance problem to get three datasets (Original, SMOTE and ADASYN datasets). Further, feature Selection methods were applied on these datasets to give three feature sets which were All Metrics (ALLM), significant features (SIGM), and uncorrelated significant features (UCORM). For each dataset, eight smells are identified, and eight different models are used, which gives us a total of (3x3x8x8) 576 Distinct Models. The predictive ability of these code-smell prediction models is compared using Accuracy, AUC, and F-Measure values. Figure \ref{conf} displays the ROC (Receiver operating characteristic) curve for three code smells (BLOB, LK and SAK). All eight models, which used significant features and trained using SMOTE sampled data, are shown in the figure. The high true positive rate in Figure \ref{conf} suggested that the trained models can predict code smell using different extracted features. Table \ref{tab21} displays the value of AUC and Accuracy for 192 out of 576 models on SMOTE Data-set. The following observations are extracted from the information present in Table \ref{tab21}: 

\begin{itemize}
    \item The accuracy of the models varies greatly, having a range from 74.52 to 96.84. However, AUC does not vary a lot with a range of .83 to .98.
	\item The mean model accuracy for SMOTE data is 80.30, and the 75th percentile is 88.14. The mean AUC value was found to be .87 and .94 being the 75th percentile. These values overall indicate high efficiency of the models developed.  
	\item The models developed using Significant features (SIGM) perform better as compared to models using Uncorrelated features (UCORM) with mean accuracy being 78.52 for SIGM and 76.52 for UCORM. 
	\item The deep learning models with eight hidden layers have the best performance as compared to models with lesser hidden layers.
\end{itemize}

\section{Comparison}
\label{sec:comparison}

\begin{figure*}[b!]
    \vspace{-0.5cm}
	\centering
	\includegraphics[width=11cm, height=3cm]{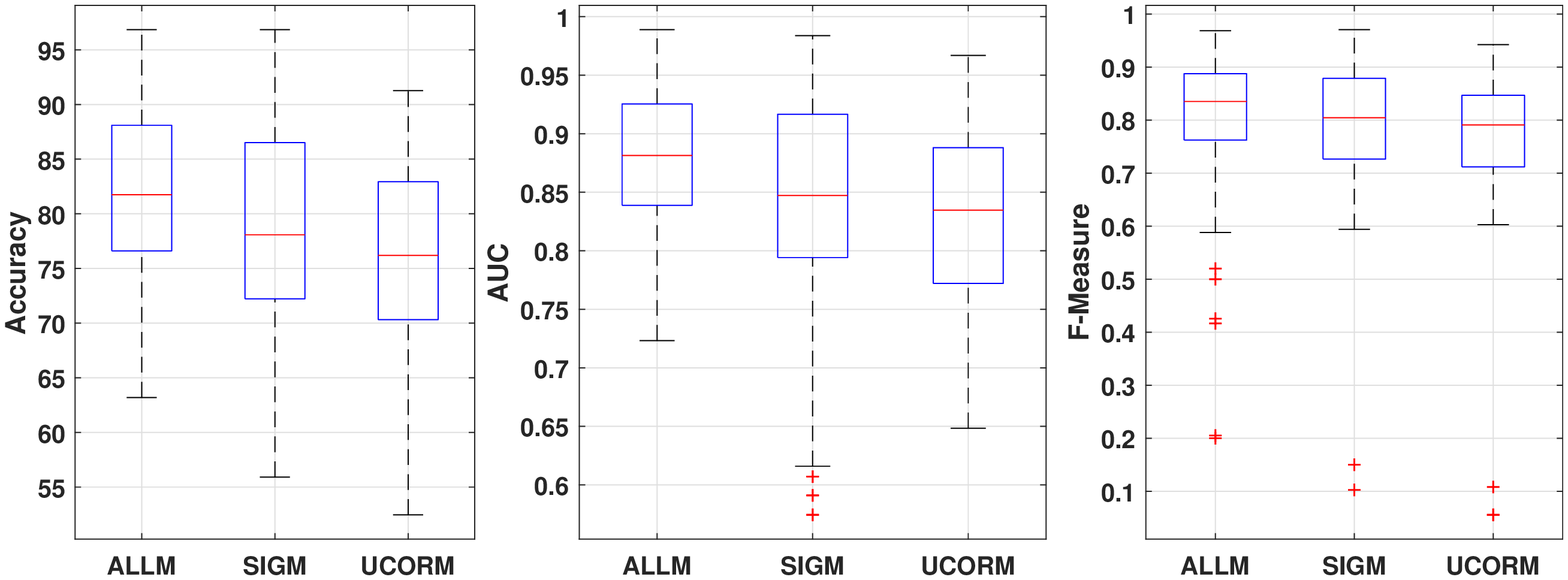}
	\caption{ Results based on Significant Metrics }
	\label{flwchrt4}
\end{figure*}

\subsection{\textbf{RQ1: Discuss the ability of selected features over the original features towards detecting Code Smell.}}

In this section, we analyze the performance of the features used within the model, obtained with the help of the Wilcoxon Sign Rank test and cross-correlation analysis. The box plots give us an easy way to visualize the predictive ability of the trained models. The models trained using different sets of features are further validated by the rank-sum test, which is used to be used for validating the considered hypothesis. In this direction, the following hypotheses are designed and tested using the rank-sum test:
\begin{itemize}
	\item \textbf{\textit{Null-Hypothesis:}} There is no significant improvement in the performance of the models trained using selected sets of features. 
	\item \textbf{\textit{Alternate hypothesis}} There is a significant improvement in the performance of the models trained using selected sets of features. 
\end{itemize}
\textbf{Comparison of different combinations of features using the box-plot diagram:}
Figure \ref{flwchrt4} shows the box-plot diagram and descriptive statistics for the performance of the trained models using three different combinations of features, namely: All Metrics (ALLM), significant metrics obtained from the Wilcoxon Sign Rank Test (SIGM), and features obtained from the Cross-Correlation Analysis (UNCORM) in terms of AUC, F-Measure, and Accuracy. The upper and lower edges of the box plot refer to the first and the third quartile value. Also, the top and the bottom line refer to the maximum and minimum value, respectively. The line inside the boxes refers to the average value of the data. In a nutshell, this diagram tells us about distribution for maximum, minimum, percentiles, and dispersion of data. 
It is evident from Figure \ref{flwchrt4} that the models trained by considering all features as an input have better ability to predict code-smell as compared to other sets of features. Figure \ref{flwchrt4}  also suggests that the models trained using significant sets of features have a better ability to predict code-smell as compared to uncorrelated sets of features. 

\textbf{Comparison of different combinations of features using Ranksum Test:}
This hypothesis has been validated at a confidence level of 95\% on the AUC value of the trained models. Hence, the considered null hypothesis is rejected if the p-value is less than 0.05, and the alternate hypothesis is rejected if the p-value is more than 0.05. Table \ref{sst} shows the results after applying the rank-sum test on the performance of the models trained using different sets of features. The results in  Table \ref{sst} suggested that the developed models using different sets of features are significantly different, i.e., the calculated p-value is smaller than 0.05 (alternate hypothesis is accepted). 

\subsection{\textbf{RQ2: Discuss the ability of Data Sampling Techniques to detect  Code Smell.}}

\begin{figure*}[b]
    \vspace{-0.5cm}
	\centering
	\includegraphics[width=11cm, height=3cm]{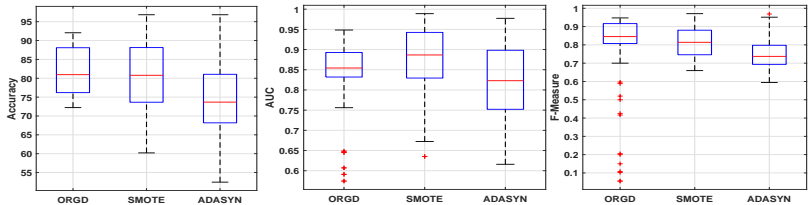}
	\caption{Result variations on Data Sampling Techniques }
	\label{flwchrt3}
\end{figure*}

In this question, we analyze the difference in performance for the models trained using datasets generated by the class imbalance techniques. As in the previous section, we make use of the same tools: box-plots, descriptive statistics, and rank-sum test of performance parameters to compare the predictive ability of the trained models using sampled data.\\
\textbf{Comparison of different samples using the box-plot diagram:}
Figure \ref{flwchrt3} shows the box-plot diagram and descriptive statistics for the performance of the trained models on original data (ORGD), SMOTE sampled data, and ADASYN sampled data in terms of AUC, F-Measure, and Accuracy. From Figure \ref{flwchrt3}, it is observed that the models trained using SMOTE techniques have better ability to predict code-smell as compared to the original data. The figure also suggested that the 25\% of the trained models on sample data have more than 0.94 AUC value. 

\begin{table*}[t!]
	\renewcommand{\thesubfigure}{\thefigure.\arabic{subfigure}}
	\centering
	\caption{Ranksum Test}
	\label{rnf}
	\subfloat[Ranksum Test: Different Model similarity\label{sst3}]{
		\renewcommand{\arraystretch}{1.1}
		\resizebox{6.0cm}{!}{
			\begin{tabular}{|l|r|r|r|r|r|r|r|r|}
			\hline
			& \multicolumn{1}{l|}{\textbf{DL1}} & \multicolumn{1}{l|}{\textbf{DL2}} & \multicolumn{1}{l|}{\textbf{DL3}} & \multicolumn{1}{l|}{\textbf{DL4}} & \multicolumn{1}{l|}{\textbf{DL5}} & \multicolumn{1}{l|}{\textbf{DL6}} & \multicolumn{1}{l|}{\textbf{DL7}} & \multicolumn{1}{l|}{\textbf{DL8}} \\ \hline
			\textbf{DL1} & 1.00                              & 0.04                              & 0.02                              & 0.01                              & 1.00                              & 0.04                              & 0.02                              & 0.01                              \\ \hline
			\textbf{DL2} & 0.04                              & 1.00                              & 0.80                              & 0.49                              & 0.04                              & 0.99                              & 0.80                              & 0.49                              \\ \hline
			\textbf{DL3} & 0.02                              & 0.80                              & 1.00                              & 0.64                              & 0.02                              & 0.80                              & 0.99                              & 0.64                              \\ \hline
			\textbf{DL4} & 0.01                              & 0.49                              & 0.64                              & 1.00                              & 0.01                              & 0.48                              & 0.64                              & 1.00                              \\ \hline
			\textbf{DL5} & 1.00                              & 0.04                              & 0.02                              & 0.01                              & 1.00                              & 0.04                              & 0.02                              & 0.01                              \\ \hline
			\textbf{DL6} & 0.04                              & 0.99                              & 0.80                              & 0.48                              & 0.04                              & 1.00                              & 0.80                              & 0.49                              \\ \hline
			\textbf{DL7} & 0.02                              & 0.80                              & 0.99                              & 0.64                              & 0.02                              & 0.80                              & 1.00                              & 0.64                              \\ \hline
			\textbf{DL8} & 0.01                              & 0.49                              & 0.64                              & 1.00                              & 0.01                              & 0.49                              & 0.64                              & 1.00                              \\ \hline
	\end{tabular}} }

	\subfloat[Different sampling methods\label{sst2}]{
		\renewcommand{\arraystretch}{1.1}
		\resizebox{4.50cm}{!}{
			\begin{tabular}{|l|r|r|r|}
			\hline
			& \multicolumn{1}{l|}{\textbf{ORGD}} & \multicolumn{1}{l|}{\textbf{SMOTE}} & \multicolumn{1}{l|}{\textbf{ADASYN}} \\ \hline
			\textbf{ORGD}   & 1.00                               & 0.00                                & 0.00                                 \\ \hline
			\textbf{SMOTE}  & 0.00                               & 1.00                                & 0.00                                 \\ \hline
			\textbf{ADASYN} & 0.00                               & 0.00                                & 1.00                                 \\ \hline
	\end{tabular}}} 
	\subfloat[Feature Combinations\label{sst}]{
		\renewcommand{\arraystretch}{1.1}
		\resizebox{4.0cm}{!}{
			\begin{tabular}{|l|c|c|c|}
			\hline
			& \multicolumn{1}{l|}{\textbf{ALLM}} & \multicolumn{1}{l|}{\textbf{SIGM}} & \multicolumn{1}{l|}{\textbf{UCORM}} \\ \hline
			\textbf{ALLM}  & 1.00                               & 0.00                               & 0.00                                \\ \hline
			\textbf{SIGM}  & 0.00                               & 1.00                               & 0.04                                \\ \hline
			\textbf{UCORM} & 0.00                               & 0.04                               & 1.00                                \\ \hline
	\end{tabular} }}
	\vspace{-1cm}
\end{table*} 

\textbf{Comparison of Different Sampling Techniques using Ranksum Test:}

The rank-sum test further validates the predictive ability models trained on different datasets. The following hypotheses are designed and tested using the rank-sum test: 
\begin{itemize}
	\item \textbf{\textit{Null-Hypothesis:}} There is no significant improvement in the performance of the models trained on balanced data. 
	\item \textbf{\textit{Alternate hypothesis}} There is a significant improvement in the performance of the models trained on balanced data.
\end{itemize}
Table \ref{sst2} shows the results after applying Ranksum test on the performance of the models trained on different datasets.  The p-value smaller than 0.05 of rank-sum test present in Table \ref{sst2} suggested that the models trained on balanced data have significant improvement in predicting different code smells. 
\vspace*{-\baselineskip}

\subsection{\textbf{RQ3: Discuss the ability of different deep learning models to detect Code Smell.}}
In this research question, we compare the performance of the code smell prediction models trained using eight different neural networks. We have used graphical analysis with the help of box-plots on Accuracy and Area Under Curve Metrics for each classifier to find the most accurate deep learning model. We have also applied the rank-sum test to validated the following hypotheses:

\begin{itemize}
	\item \textbf{\textit{Null-Hypothesis:}} There is no significant improvement in the performance of the models while increasing the number of hidden layers. 
	\item \textbf{\textit{Alternate hypothesis}} There is a significant improvement in the performance of the models while increasing the number of hidden layers. 
\end{itemize}

\begin{figure*}[b!]
    \vspace{-0.5cm}
	\centering
	\includegraphics[width=11cm, height=3cm]{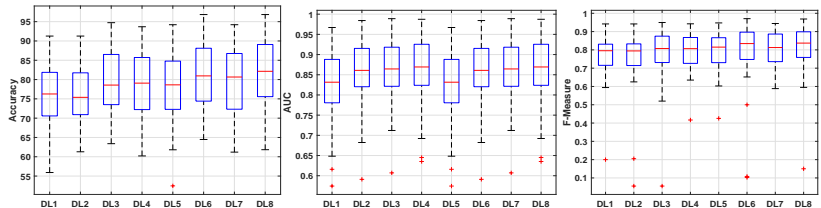}
	\caption{Results for Different Deep Learning Models}
	\label{flwchrt2}
\end{figure*}

\textbf{Comparison of different Classifiers using Descriptive Statistics and the box-plot diagram:}
Figure \ref{flwchrt2} shows the box-plot diagram for the performance of the trained models using eight deep learning models in terms of AUC, F-Measure, and Accuracy. We see that, as the number of hidden layer increases, a corresponding increase in performance in the prediction of the models. However, from Figure \ref{flwchrt2} we can a slight dip in performance in architecture with seven hidden layers (Mean Values: 79.39 accuracy, .86 AUC and .81 F-Measure respectively) as compared to six and eight hidden layers (Mean Values: 81.71 and 81.16 accuracy, .86 and .86 AUC and .81 and .82 F-Measure respectively).

\textbf{Comparison of Different Classifying Techniques using Ranksum Test:}
The above-considered hypothesis has been validated at a confidence level of 95\% on the AUC value of the trained models. Hence, the null hypothesis is rejected if the p-value is less than 0.05, and the alternate hypothesis is rejected if the p-value is more than 0.05. Table \ref{sst3} shows the results after applying the rank-sum test on the performance of the models. The results in Table \ref{sst3} suggested that there is no significant improvement in the performance of the models while increasing the number of hidden layers, i.e., the calculated p-value for most of the pairs are more than 0.05 (alternate hypothesis is rejected).

\section{Conclusion}
\label{sec:conclusion}
This paper presents the empirical analysis on code smell prediction models developed using data sampling, features selection and neural networks. These models are validated using 5-fold cross-validation, and the prediction ability of these models are compared using three different performance parameters which were AUC, Accuracy, and F-Measure. 

Our primary conclusion is that an increase in the number of hidden layer did not lead to a monotonic increase in performance contrary to prior expectation. Furthermore, we see diminishing returns in the performance increase with the addition of a hidden layer. We also observe that the models with eight hidden layers performs the best (higher accuracy, AUC and F-Measure) as compared to other models. The AUC, Accuracy, and F-Measure values of the trained models also suggest that the models trained on balanced data perform better than models developed on original data. The rank-sum test also indicates that the models trained using balanced data have a significant improvement in code smell prediction ability. The results of the rank-sum test also indicate that there is no significant improvement in the performance of the models while increasing the number of hidden layers. 
\vspace{-.5cm}
\bibliographystyle{unsrt}
\bibliography{bibliography}

\end{document}